\documentclass[5p,fleqn]{elsarticle}

\usepackage{graphicx, amsmath}
\begin{document}
\begin{frontmatter}

\title{Non-scalar Contribution of Potential in Crystals}

\author{
H. Uchiyama}
\address{
Japan Synchrotron Radiation Research Institute, Koto 1-1-1, Sayo, Hyogo 679-5198, Japan.
}%

\date{\today}
\begin{abstract}
Ewald-parameter dependence of Coulomb interaction in ionic crystals was studied using a point-charge model. 
In the presence of the long-range interaction, the ion configuration breaks spherical symmetry of local potential and charge at each ion site, and gives non-scalar contributions to them. 
This non-scalar potential has similar effects to Heisenberg interaction, while is intrinsically distinct from conventional multipole expansions of the scalar potential.
Symmetry and magnitude of the scalar and non-scalar potentials are similar for most materials despite the different definitions, but one exception can be seen in parent materials of hole-doped high-$T_{\mathrm{c}}$ cuprates. 
\end{abstract}
\end{frontmatter}

\section{Introduction}
In crystal-field theory, anisotropic Coulomb potential causes deviation of wavefunctions from spherical symmetry around a site of an ionic crystal.\cite{Hutchings} It is well known that $d$ electrons in the octahedral field produced by six surrounding anions split into $t_{2g}$ and $e_g$ orbitals with different energies. The anisotropic potential is described as a multipole expansion, a scalar function of the position. The potential is originally composed of contributions from the nearest neighboring ions with high symmetry, or cubic groups. More general consideration of the anisotropic potential, which includes contributions beyond the neighboring ions with lower symmetry, is discussed using the Ewald method.\cite{Nijboer,Tosi,Rudge} However, Ref.~\cite{Nijboer} indicates that this method has conditional convergence for the multipole expansions. The treatment beyond the Ewald method is required for the absolute convergence, and this treatment for charge density is applied to the first principle calculations.\cite{ Bertaut,Weinert} 
 Some reports mention that the conditional convergence is caused by the shape of the crystal.\cite{Smith}

Classical Heisenberg interaction discusses another deviation from spherical symmetry. This interaction is controlled by spin orientations at the nearest neighboring sites, similar to the crystal field theory, but describes local rotational symmetry exactly at the site, in contrast to the anisotropy above. Given the fact that rotational symmetry at a certain point in general includes contributions from all ions in the crystal, there is some possibility of unknown interaction at the site, which is not described by the Heisenberg interaction only. 

In this report, Coulomb interaction in crystals with infinite periodicity were reinvestigated in the point-charge model, based on the Ewald method. It turns out that, in the presence of the long-range interaction, the Coulomb potential has another, non-scalar, contribution, which breaks the spherical symmetry at the site, in addition to the scalar contribution. This dual aspects of the potential may cause the conditional convergence of the Ewald method. The non-scalar contribution, which has the same deviation of the spherical symmetry as the Heisenberg interaction, is caused by the ion configuration of the infinite lattice, and coexists with distortion of the charge from the original spherical symmetry. This non-scalar contributions neither violate Poisson's equation nor affect scalar potential and charge. Furthermore, it can be defined both at magnetic and non-magnetic ions, in contrast to the Heisenberg interaction, and be considered as anisotropy of local relative permittivity in the framework of the scalar-potential field.

As specific examples, anisotropic potentials and charges of a $d$-ion in a NaCl-type structure and ions in ZnO, ZnS (zinc blende), CaF$_2$, TiO$_2$ (rutile), SrTiO$_3$, La$_2$CuO$_4$, Nd$_2$CuO$_4$, and HgBa$_2$CuO$_4$ were calculated. Though these two (scalar and non-scalar) potentials are defined differently, they have the same symmetries and similar magnitudes at the ion sites in SrTiO$_3$ and the $d$-ion site in the NaCl-type structure. The similar features are also observed in more complicated materials, such as ZnO, ZnS, CaF$_2$, TiO$_2$, and Nd$_2$CuO$_4$. However, these potentials have completely different symmetries at the O sites in the CuO$_2$ planes of La$_2$CuO$_4$ and HgBa$_2$CuO$_4$, parent materials of hole-doping high-$T_{\mathrm{c}}$ cuprates. This difference may have some connection with the hole-doped superconductivity. 

\section{Method}
In the Ewald method, the Coulomb potential induced by the surrounding point charges at an $i$-th ion site ($\mathbf{r}_i$) is expressed as follows, using a parameter $\xi$;
\begin{equation}
 V(\mathbf{r}_i) =\sum\limits_{j \ne i} {f(\xi)}+ \sum\limits_j {\sum\limits_{k \ne 0} {g(\xi)} }- \frac{ Z_ie}{2\pi^{3/2}\varepsilon_0 }\xi,
\end{equation}
when the crystal has charge neutrality. 
Here, $\varepsilon_0$ is the vacuum permittivity, $f(\xi)$ is the contribution from real space:
\begin{equation}
f(\xi)= \frac{Z_je}{4\pi\varepsilon_0 \left| \mathbf{r}_j - \mathbf{r}_i \right|} \mathrm{erfc}\left( \left| {\mathbf{r}_j - \mathbf{r}_i} \right|\xi \right),
\end{equation}
and $g(\xi)$ is the contribution from reciprocal space:
\begin{equation}
g(\xi)= \frac{Z_je}{4 \pi^2 k^2 \varepsilon_0 v}e^{ -\pi ^2 k^2/\xi ^2}e^{2\pi i \mathbf{k} \cdot (\mathbf{r}_j - \mathbf{r}_i) } ,
 \end{equation}
where $v$ is the volume of the unit cell.

When $\xi$ is small enough ($\xi \to 0$), $g(\xi)$ approaches zero, and Eq.~(1) becomes
\begin{equation}
V(\mathbf{r}_i) + \frac{Z_ie}{2\pi^{3/2}\varepsilon_0 }\xi \sim \sum\limits_{j \ne i} {f}.
\end{equation}
This equation indicates that the potential and charge at the $i$-th site are determined by the surrounding ions ($j\ne i$) for small $\xi$. Furthermore, because the right side of Eq.~(4) is not a summation of distribution with spherical symmetry, the left side is also expected to lose the symmetry and deviate from scalars.

When the spherical symmetry of the local potential and charge are broken at $\mathbf{r}_i$, Eq.~(1) can be expanded using associated Legendre polynomials ($P_l^m$) in spherical polar coordinates $(\theta,\varphi)$;
\begin{equation}
V_i^{(n)}(\theta,\varphi) + \frac{Z_i^{(n)}(\theta,\varphi) e}{2\pi^{3/2}\varepsilon_0 }\xi = \sum\limits_{l = 0}^n k_{nl} L_i^{(l)}(\xi,\theta,\varphi),
\end{equation}
for the $n$-th order expansion. Here,
\begin{eqnarray}
&&L_i^{(l)}=a_i^{l0}P_l^0(\cos \theta ) \nonumber\\
&&+\sum\limits_{m = 1}^l {\left[ {a_i^{lm}P_l^m(\cos \theta )\cos m\varphi + b_i^{lm}P_l^m(\cos \theta )\sin m\varphi } \right]}, \nonumber
\end{eqnarray}
\begin{eqnarray*}
a_i^{lm}(\xi) = C_{lm}\sum\limits_{j \ne i} {fP_l^m(\cos \theta )\cos m\varphi } \\
+ C_{lm}\sum\limits_j {\sum\limits_{k \ne 0} {g P_l^m(\cos \theta )\cos m\varphi } },
\end{eqnarray*}
and
\begin{eqnarray}
b_i^{lm}(\xi) = C_{lm}\sum\limits_{j \ne i} {fP_l^m(\cos \theta )\sin m\varphi } \nonumber \\
+ C_{lm}\sum\limits_j {\sum\limits_{k \ne 0} {g P_l^m(\cos \theta )\sin m\varphi } } .
\end{eqnarray}
The coefficient $C_{lm}$ may be given as
\begin{eqnarray}
&&C_{lm} =(-1)^m \frac{2(l- m)!}{(l+m)!}C_{l0} \;\; \; \; (m \ge 1),\nonumber\\
&&C_{l0} = \frac{l!}{(2l - 1)!!}.
\end{eqnarray}
The $n$-th order Legendre expansion of the anisotropic potential is alternatively expressed as the $n$-fold tensor product of the unit vector, ($\sin\theta\cos\varphi$, $\sin\theta\sin\varphi$, $\cos\theta$),\cite{TMO} and hence the coefficient $k_{nl}$ should satisfy the following equation,
\begin{equation}
\cos ^n\theta + \cos ^{n - 1}\theta = \sum\limits_{l = 0}^n {k_{nl}C_{l0}P_l^0(\cos \theta )},
\end{equation}
for even $n(\ge2)$ ($k_{00}=1$ for $n=0$). It should be noted that the charges in Eqs.~(2) and (3) ($Z_j$) should also deviate from scalars, because Eq.~(5) indicates deviation from the spherical (isotropic) charge at $\mathbf{r}_i$. However, this further deviation is not considered here for simplicity. 

 \begin{figure}
\includegraphics[width=1\linewidth,clip=]{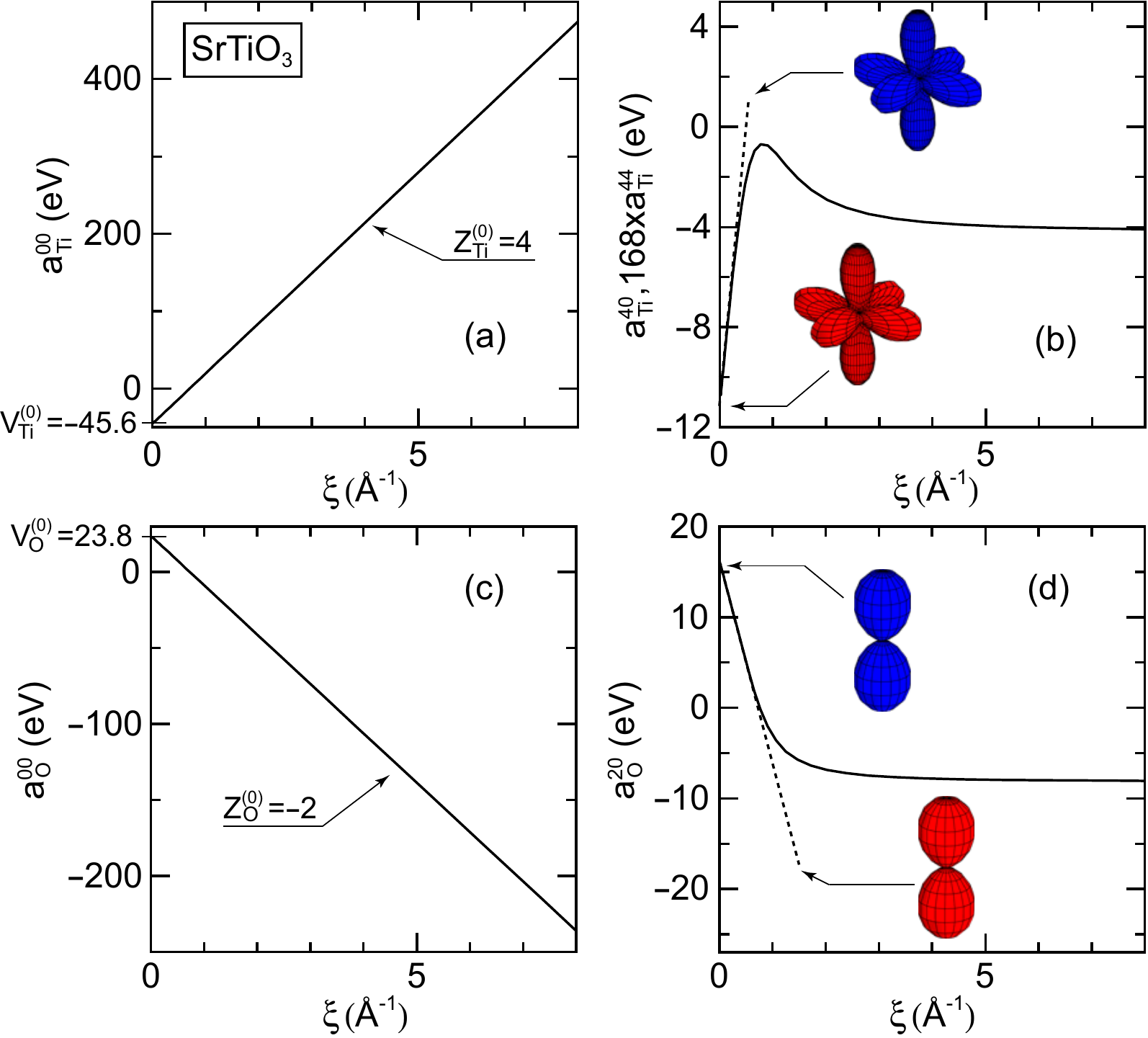}
\caption{\label{fig1} Non-zero contributions of $L_i^{(l)}$ at the Ti and O sites in cubic SrTiO$_3$. (a, c) The scalar terms ($l=0$) at Ti (a) and O (c). (b) The 4th order contribution ($l=4$) at the Ti site. (d) The 2nd order contribution ($l=2$) at the O site. Inset figures in (b) and (d) indicate the anisotropic potentials and charges, estimated by the dotted lines, or linear fits for small $\xi$. The blue drawings have positive values (deviation from the minimum), while the red drawings have negative values (deviation from the maximum).}
\end{figure}

For instance, the Legendre expansions at Ti and O sites in a cubic perovskite, SrTiO$_3$ ($a$=3.905 \AA), are considered (Fig.~1). In the crystal, scalar point charges (Sr$^{2+}$, Ti$^{4+}$, and O$^{2-}$) are assumed, which behave as ions. Figs.~1(a) and (c) show $a_{\mathrm{Ti}}^{00}(\xi)$ and $a_{\mathrm{O}}^{00}(\xi)$, or $\xi$ dependence of the scalar (isotropic) terms ($l=0$) at the Ti and O sites. Eq.~(5) indicates that the isotropic potentials are given by the $a_i^{00}$-intercepts ($V_{\mathrm{O}}^{(0)}$ and $V_{\mathrm{Ti}}^{(0)}$), and that the isotropic charges are given by the slopes ($Z_{\mathrm{O}}^{(0)}$ and $Z_{\mathrm{Ti}}^{(0)}$). The former exactly corresponds to the conventional scalar potentials ($V_{\mathrm{O}}^{(0)}$=23.8 eV and $V_{\mathrm{Ti}}^{(0)}$=$-$45.6 eV) and the latter gives the self charges at $\mathbf{r}_i$ ($Z_{\mathrm{O}}^{(0)}$=$-$2 and $Z_{\mathrm{Ti}}^{(0)}$=+4). These values are unchanged in an arbitrary region of $\xi$. 

At the Ti site, though the anisotropic contributions of $l$=1, 2, and 3 are zero, non-zero contribution appears at the 4th order ($L_{\mathrm{Ti}}^{(4)}\ne 0$). When the Ti-O bonds are directed to the Cartesian coordinates, $a^{40}_{\mathrm{Ti}}$ and $a^{44}_{\mathrm{Ti}}$ become non-zero, as seen in Fig.~1(b). In contrast to the isotropic contribution ($l=0$, Fig.~1(a)), these coefficients have $\xi$ dependence. For small $\xi$, we can estimate the non-scalar potential and charge using the intercepts and slopes of the dotted lines in Fig.~1(b). One of the inset figures (blue) in Fig.~1(b) suggests positive charge anisotropy determined by the slopes of $a^{40}_{\mathrm{Ti}}$ and $a^{44}_{\mathrm{Ti}}$, and the other (red) suggests negative potential anisotropy obtained by the intercepts. Both the anisotropic potential and charge have $O_h$ symmetry, but the signs are opposite. These features are similar to the isotropic contributions, where the isotropic (scalar) charge and potential ($Z_{\mathrm{Ti}}^{(0)}$ and $V_{\mathrm{Ti}}^{(0)}$) have the same (spherical) symmetry with the opposite signs, as seen in Fig.~1(a). This similarity validates the existence of the non-scalar (anisotropic) charge and potential at small $\xi$. 

At the O site, non-zero contribution appears at the 2nd order ($L_{\mathrm{O}}^{(2)}\ne 0$); when the $z$-axis is taken along the Ti-O bond direction, $a^{20}_{\mathrm{O}}$ becomes non-zero as shown in Fig.~1(d). Similar to the Ti site, the anisotropic potential and charge at O can be defined (only) for small $\xi$; the slope and intercept of $a^{20}_{\mathrm{O}}$ give negative charge anisotropy (red) and positive potential anisotropy (blue), as shown in the insets of Fig.~1(d). Again, the potential and charge have the same symmetry with the opposite signs. 

These non-scalar contributions of the potential and charge have several features. First, these do not violate the Poisson's equation around the $i$-th ion; only the scalar term of the charge ($Z_i^{(0)}$) at $\mathbf{r}_i$ contributes to the equation, because Eq.~(5) satisfies the following equation,
\begin{equation}
(n+1)\iint {\mathop {\lim }\limits_{\xi \to 0} Z_i^{(n)}\sin\theta d\theta d\varphi }= 4\pi Z_i^{(0)},
\end{equation}
for even $n$. Secondly, the non-scalar potential and charge reflect rotational symmetry at the $i$-th site, because the Legendre expansion in Eq.~(5) is chiefly affected by the surrounding ions ($f(\xi)$) for small $\xi$. Finally, the non-scalar potential and charge for small $\xi$ suggest some relation to the long-range Coulomb interaction, because $\xi$ has the dimension of $1/r$. These features suggest spherical symmetry breaking of the potential and charge at $\mathbf{r}_i$, owing to the long-range Coulomb interaction. 

In the conventional crystal-field theory, on the other hand, anisotropic Coulomb potential caused by the neighboring ions around $\mathbf{r}_i$ is expressed as a scalar function of the position. When contribution of the infinite periodicity (beyond the neighboring ions) is included in the theory, the scalar potential at $\mathbf{r}_s+\mathbf{r}_i$ is given by
\begin{equation}
V_i(\mathbf{r}_s) =\sum\limits_{j \ne i} f (\xi ,\mathbf{r}_s) + \sum\limits_j {\sum\limits_{k \ne 0} {g(\xi ,\mathbf{r}_s})}- \frac{ Z_i e}{ 4\pi\varepsilon_0} \frac{\mathrm{erf}(\xi r_s)}{r_s}.
 \end{equation}
Here,
\begin{equation}
 f (\xi ,\mathbf{r}_s)= \frac{Z_je}{4\pi\varepsilon_0 \left| \mathbf{r}_j - \mathbf{r}_i-\mathbf{r}_s \right|} \mathrm{erfc}\left( \left| {\mathbf{r}_j - \mathbf{r}_i-\mathbf{r}_s} \right|\xi \right),\nonumber
\end{equation}
and
\begin{equation}
g(\xi,\mathbf{r}_s)= \frac{Z_je}{4 \pi^2 k^2 \varepsilon_0 v}e^{ -\pi ^2 k^2/\xi ^2}e^{2\pi i \mathbf{k} \cdot (\mathbf{r}_j - \mathbf{r}_i-\mathbf{r}_s) } .
 \end{equation}
 The equations become identical with Eqs.~(1)--(3) in the limit of $r_s\to 0$, and are essentially the same as those in Ref.\cite{Nijboer}. Note, there is no $\xi$ dependence on this scalar potential. 
 
 In spherical polar coordinates of $\mathbf{r}_s$, $(r_s,\theta ,\varphi )$, Eq.~(10) is expressed as a multipole expansion;
\begin{equation}
V_{i}^{(n)}(r_s,\theta ,\varphi ) = \sum\limits_{l = 0}^n {{r_s^l}} {L'}_i^{(l)}(\theta ,\varphi ) ,
\end{equation}
when the contributions to the $n$-th order are taken into account. The coefficient ${L'}_i^{(l)}$ is given as
\begin{eqnarray}
&&{L'}_i^{(l)}={a'}_i^{l0}P_l^0(\cos \theta ) \nonumber\\
&&+\sum\limits_{m = 1}^l {\left[ {{a'}_i^{lm}P_l^m(\cos \theta )\cos m\varphi + {b'}_i^{lm}P_l^m(\cos \theta )\sin m\varphi } \right]}, \nonumber \\&&
\end{eqnarray}
similar to $L_i^{(l)}$ in Eq.~(5).
When Eq.~(12) is estimated using the results of Eq.~(10), ${L'}_{i}^{(l)}$ is determined uniquely (independent of $\xi$), in contrast to the direct estimation of Eq.~(12) reported in Ref.\cite{Nijboer}. This ${L'}_{i}^{(l)}$ is essentially distinct from the non-scalar contribution $L_{i}^{(l)}$; ${L'}_{i}^{(l)}$ specifies the potential only, while $L_{i}^{(l)}$ determines both the potential and charge. 

 In short, the anisotropic Coulomb potential can be defined in two ways in the framework of the point-charge (ionic) model. One is the multipole expansion of the scalar potential around $\mathbf{r}_i$, which is given by Eqs.~(10) and (12) (hereafter it is called $V^{\mathrm{S},(n)}_i(r_s,\theta,\varphi)$ for classification). This scalar potential is conventional and corresponds to the crystal-field theory for the infinite lattice. The other is obtained from the non-scalar contribution of Eq.~(5) at the $i$-th site (in the limit of $\xi\to 0$, $V^{\mathrm{N},(n)}_i(\theta,\varphi)$). The potential is accompanied by the non-scalar charge ($Z^{\mathrm{N},(n)}_i(\theta,\varphi)$). Both scalar and non-scalar potentials have absolute convergence in the Ewald method.

The latter non-scalar potential corresponds to the rotational symmetry at the site.
Given the fact that the non-scalar contribution ($L_{i}^{(l)}$, $l>0$) does not affect the scalar potential and charge, the scalar and non-scalar potential should work independently. Furthermore, this has similar contribution to the Heisenberg interaction (Note, the non-scalar potential works even at the non-magnetic ions), because both of them define anisotropy at the site. The non-scalar potential has orientation dependence at the site, which is understood as a anisotropic relative permittivity in the framework of the conventional scalar potential, while it is different from the dipole interaction caused by the scalar potential.

\section{Results and Discussion}

In order to discuss the anisotropy, anisotropic potential at the $i$-th site, $\Delta V_i^{{\mathrm{S}},(n)}(r_s,\theta,\varphi)$, for $ V_i^{{\mathrm{S}},(n)}$ is defined as
\begin{eqnarray}
\Delta V_i^{\mathrm{S},(n)}=V_i^{\mathrm{S},(n)}- V^{\mathrm{S},(n)}_{i,min} \;\; \; \; (V_i^{\mathrm{S}, (0)} > 0) \nonumber\\
= V_i^{\mathrm{S},(n)}-V^{\mathrm{S},(n)}_{i, Max} \;\; \; \; (V_i^{\mathrm{S}, (0)} < 0),
\end{eqnarray}
where $V^{\mathrm{S},(n)}_{i, Max}$ ($V^{\mathrm{S},(n)}_{i, min}$) is the maximum (minimum) value of $V^{\mathrm{S},(n)}_{i}(r_s,\theta,\varphi)$. In the same way, $\Delta V_i^{\mathrm{N},(n)}(\theta,\varphi)$ and $\Delta Z_i^{\mathrm{N},(n)} (\theta,\varphi)$ are defined for $V_i^{\mathrm{N},(n)}$ and $Z_i^{\mathrm{N},(n)}$. Through this manuscript, the negative anisotropic potential and charge are depicted as red, while the positive ones are displayed in blue. In calculating $\Delta V^{\mathrm{S},(n)}_i$, $r_s$=1.3 \AA\ is used in common. This $r_s$ is about two-thirds of a distance to the adjacent ion for the ions studied. 
 
Figure 2 shows anisotropic charges and potentials at the Ti and O sites in a TiO$_2$ plane of cubic SrTiO$_3$. In this Figure, the contributions to the 2nd order ($n$=2) are taken into account at the O site, and those to the 4th order ($n$=4) are taken at the Ti site, because Clebsch-Gordan coefficients expect $l\le2$ for a $p$-orbital and $l\le 4$ for a $d$-orbital. Figures 2(a) and (b) show the non-scalar charge and potential (which summarize the discussions in the previous section), while Figure 2(c) displays the multipole expansions of the scalar potentials. Figure 2(a) suggests (electron-like) negative anisotropic charge at the O site (red, $\Delta Z_{\mathrm{O}}^{\mathrm{N},(2)}$) and (hole-like) positive anisotropic charge at the Ti site (blue, $\Delta Z_{\mathrm{Ti}}^{\mathrm{N},(4)}$), while Figures 2(b) and (c) indicate positive anisotropic potential at O (blue, $\Delta V_{\mathrm{O}}^{\mathrm{N},(2)}$ and $\Delta V_{\mathrm{O}}^{\mathrm{S},(2)}$) and negative anisotropic potential at Ti (red, $\Delta V_{\mathrm{Ti}}^{\mathrm{N},(4)}$ and $\Delta V_{\mathrm{Ti}}^{\mathrm{S},(4)}$). The non-scalar potentials at the Ti and O sites (Fig.~2(a)) have the same symmetries as the respective charges (Fig.~2(b)) with the opposite signs, as mentioned in Sec.~II. Furthermore, the non-scalar potentials at Ti and O ($\Delta V_{\mathrm{Ti}}^{\mathrm{N},(4)}$ and $\Delta V_{\mathrm{O}}^{\mathrm{N},(2)}$) in Fig.~2(b) have the same symmetries as the respective scalar potentials ($\Delta V_{\mathrm{Ti}}^{\mathrm{S},(4)}$ and $\Delta V_{\mathrm{O}}^{\mathrm{S},(2)}$) in Fig.~2(c). At $r_s$=1.3 \AA, these scalar and non-scalar potentials also have similar size.

 \begin{figure}
\includegraphics[width=0.8\linewidth,clip=]{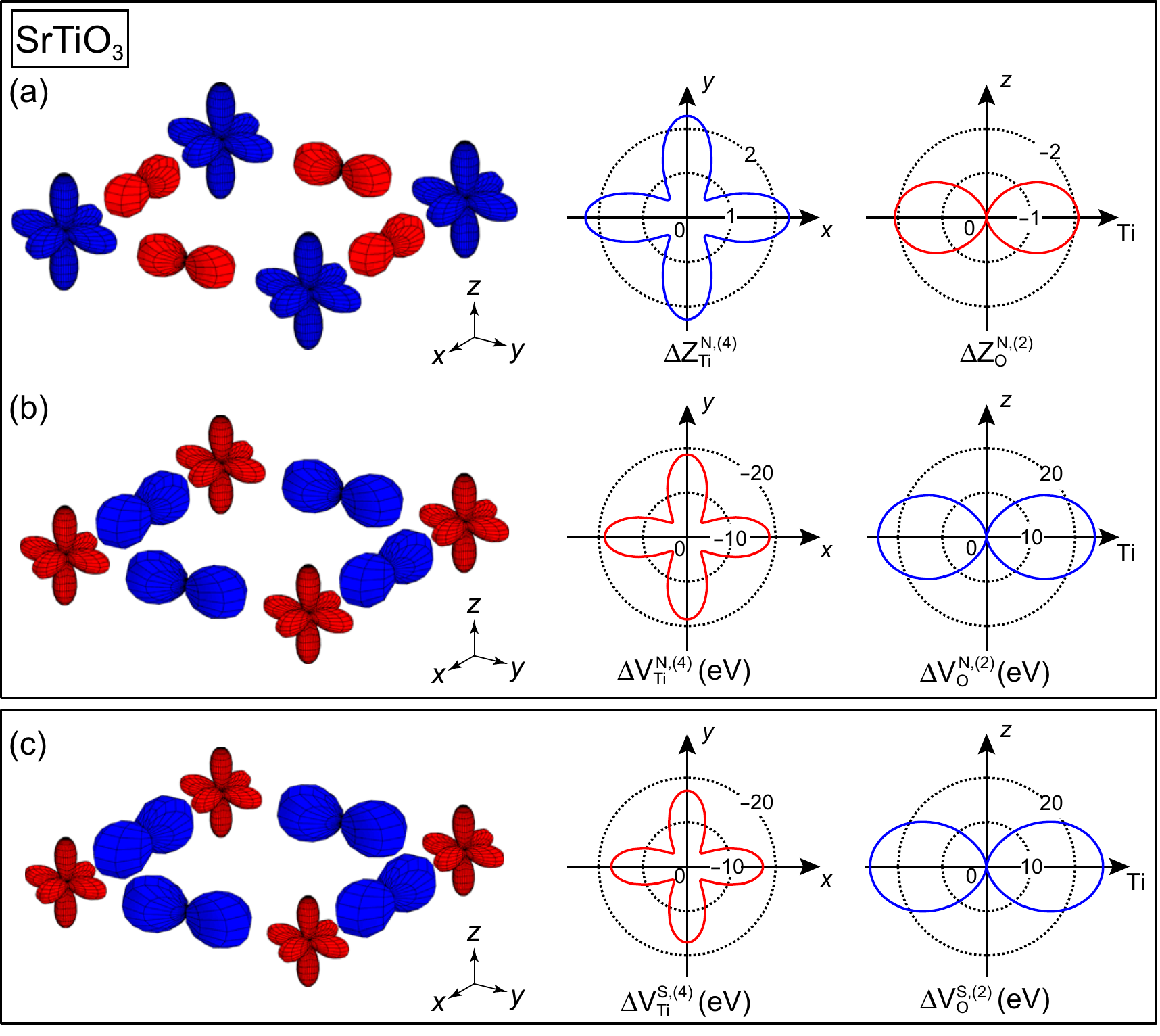}
\caption{\label{fig2} Anisotropic charges and potentials at the Ti and O sites of cubic SrTiO$_3$. (a, b) Non-scalar contributions of the charges (a) and potentials (b). (c) Multipole expansions of the scalar potentials. For each Figure, the left drawing shows the overhead view of the TiO$_2$ plane with the axes of the Cartesian coordinates. The center and right graphs indicate the anisotropy in the $xy$ and $zb_{\mathrm{TiO}}$ planes ($b_{\mathrm{TiO}}$ is the Ti-O bond direction) at the Ti and O sites, respectively.}
\end{figure}

Next, anisotropic behavior at a $d$-ion site in the NaCl-type structure is considered. It is assumed that the distance to the adjacent ion is $a$ and the charges of the ions are $\pm Z$. 
There are no anisotropic contributions of $l$=1, 2, and 3 for two anisotropic potentials and anisotropic charge. At the $d$-ion site ($n=4$), the non-scalar potential, $\Delta V_{d}^{\mathrm{N},(4)}$, and the multipole expansion of the scalar potential, $\Delta V_{d}^{\mathrm{S},(4)}$, are given by
\begin{eqnarray}
\Delta V_d^{\mathrm{N},(4)}=-0.697\frac{1}{4\pi\varepsilon_0 a}C_d^{(4)}Z \nonumber\\
\Delta V_d^{\mathrm{S},(4)}=-3.58\frac{r_s^4}{4\pi\varepsilon_0 a^5}C_d^{(4)}Z,
\end{eqnarray}
when the Cartesian coordinates are taken along the bond directions.
Here, the coefficient is given as
\begin{equation}
C_d^{(4)}=P_4^0(\cos\theta)+\frac{1}{168}P_4^4(\cos\theta)\cos4\varphi+\frac{2}{3}.
\end{equation}
These anisotropic potentials ($\Delta V_d^{\mathrm{N},(4)}$ and $\Delta V_d^{\mathrm{S},(4)}$) have $O_h$ symmetry, the same as that at Ti in SrTiO$_3$ (Fig.~2). Moreover, they have the same value at $r_s \sim 2a/3$. The anisotropic non-scalar charge, $\Delta Z_{d}^{\mathrm{N},(4)}$ (positive), has the same $O_h$ symmetry as $\Delta V_{d}^{\mathrm{N},(4)}$ and $\Delta V_{d}^{\mathrm{S},(4)}$ (negative) with the opposite sign. The anisotropic potentials and charge suggests $e_g$-like hole distribution. (In terms of the scalar potential, it agrees with the crystal-field theory, where $t_{2g}$ orbitals are stable). Given the fact that these potentials are independently defined, the non-scalar potential is supposed to give additional effects to the contribution expected by the scalar potential for stabilization of $t_{2g}$. 

As a slightly complicated case, rutile (TiO$_2$) is considered. The lattice and charge parameters are given as
$a$=4.594 \AA, $c$=2.958 \AA, $u$=0.3053, Ti$^{4+}$, and O$^{2-}$. At the O site, higher order ($n> 2$) contributions should be included, owing to the $sp$-hybridization. Figure 3, hence, displays the 4th-order contributions ($n$=4) both at the Ti and O site. Following the style of Fig.~2 for SrTiO$_3$, the non-scalar charge and potential are shown in Figs.~3(a) and (b), while the multipole expansions of the scalar potentials are displayed in Fig.~3(c). 

\begin{figure}
\includegraphics[width=1\linewidth,clip=]{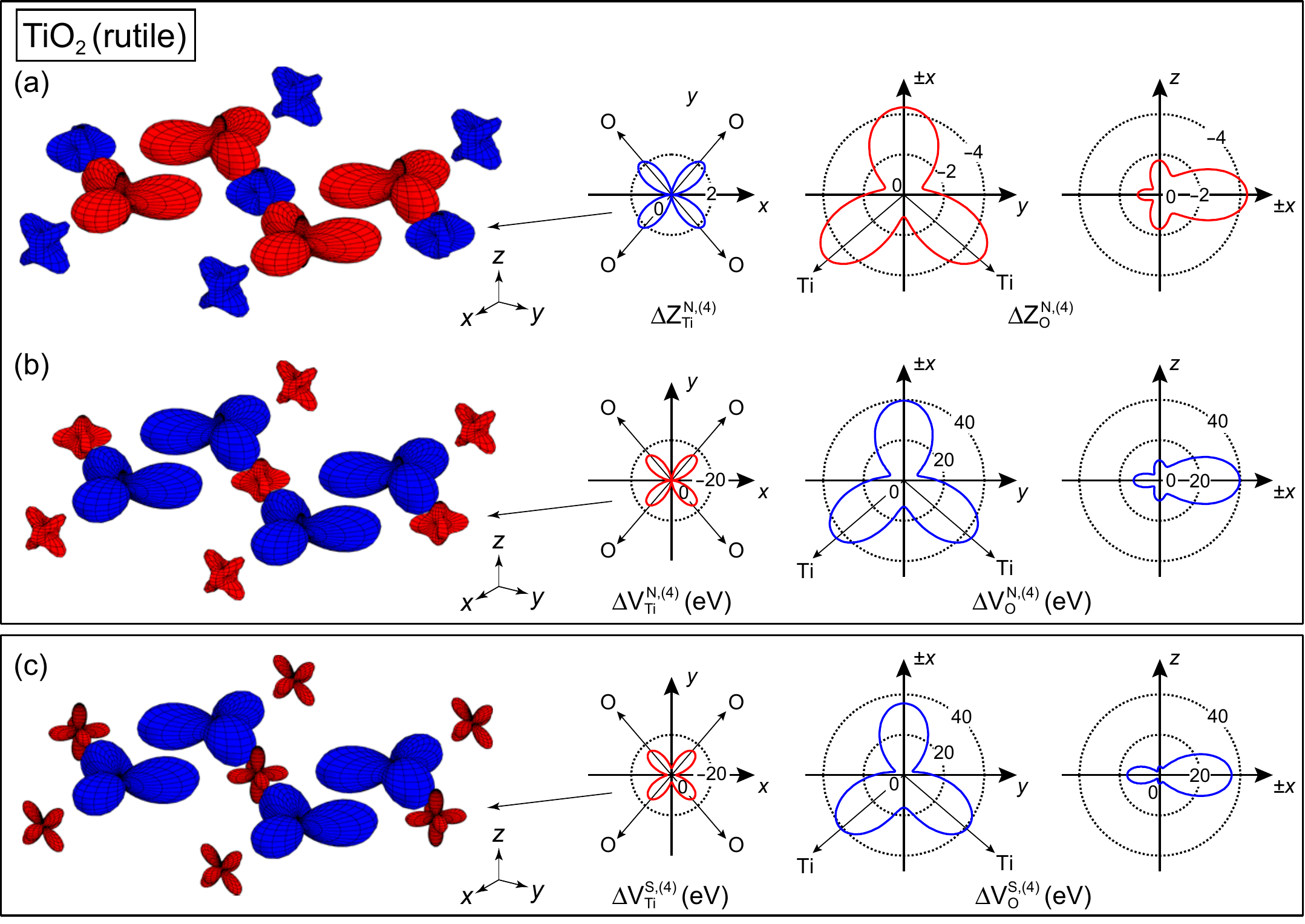}
\caption{\label{fig3}Anisotropic charges and potentials at the Ti and O sites of TiO$_2$ (rutile). (a, b) Non-scalar contributions of the charges (a) and potentials (b). (c) Multipole expansions of the scalar potentials. For each Figure, the left drawing shows the overhead view of the (110) plane with the axes of the Cartesian coordinates. The center graph indicates the anisotropy in the $xy$ plane at one of the Ti sites (indicated by the arrow). The right two graphs show the anisotropy in the $xy$ and $xz$ plane at the O site.}
\end{figure}

Figures 3(a) and (b) indicate that the non-scalar charges at the Ti and O sites ($\Delta Z_{\mathrm{Ti}}^{\mathrm{N},(4)}$ and $\Delta Z_{\mathrm{O}}^{\mathrm{N},(4)}$) are similar in shape to the respective non-scalar potentials ($\Delta V_{\mathrm{Ti}}^{\mathrm{N},(4)}$ and $\Delta V_{\mathrm{O}}^{\mathrm{N},(4)}$) with the opposite signs. $\Delta V_{\mathrm{Ti}}^{\mathrm{N},(4)}$ and $\Delta V_{\mathrm{O}}^{\mathrm{N},(4)}$ in Fig.~3(b) are similar in shape and size to $\Delta V_{\mathrm{Ti}}^{\mathrm{S},(4)}$ and $\Delta V_{\mathrm{O}}^{\mathrm{S},(4)}$ in Fig.~3(c), respectively. The obtained anisotropy at the O site ($\Delta Z_{\mathrm{O}}^{\mathrm{N},(4)}$, $\Delta V_{\mathrm{O}}^{\mathrm{N},(4)}$, and $\Delta V_{\mathrm{O}}^{\mathrm{S},(4)}$) is close to the Wannier functions determined by the first principle calculations, that indicate the $sp^2$-like orbitals along the Ti-O bonds and the $p_z$-like orbital perpendicular to the (110) plane.\cite{TiO2band} In other words, the present anisotropic potentials and charge reflect the wavefunctions, in spite of the simple point-charge model. Strictly speaking, these scalar and non-scalar contributions in Fig.~3 do not have the same symmetries, in contrast to SrTiO$_3$. For example, the shapes at the Ti site are slightly different from each other, even though they are close to $O_h$ symmetry. Moreover, the $p_z$-like component in $\Delta V_{\mathrm{O}}^{\mathrm{S},(4)}$ (the right graph in Fig.~3(c)) is considerably suppressed more than those in $\Delta Z_{\mathrm{O}}^{\mathrm{N},(4)}$ and $\Delta V_{\mathrm{O}}^{\mathrm{N},(4)}$ (those in Figs.~3(a) and (b)). As for the latter, it should be noted that the $p_z$-like component in $ \Delta V_{\mathrm{O}}^{\mathrm{S},(4)}$ is more enhanced at larger $r_s$.

In the discussion so far, the anisotropic potentials and charges share two common features. One is the similarity in shape between $\Delta V_i^{\mathrm{N},(n)}$ and $\Delta Z_i^{\mathrm{N},(n)}$, though their signs are opposite. The other is the similarity in shape and size between $\Delta V_i^{\mathrm{N},(n)}$ and $\Delta V_i^{\mathrm{S},(n)}$ (Properly speaking, the similarity in size is applicable only when $r_s$ is about two-thirds of the distance to the adjacent ion (Figs.~2, 3 and Eq.~(15)). However, if the ion at $\mathbf{r}_i$ has an effective radius, these potentials are expected to be of a similar order of magnitude). Both features are further confirmed in other materials, at anion sites in CaF$_2$, ZnO, ZnS (zinc blende) (for $n$=4, not shown). Each non-scalar charge $\Delta Z_i^{\mathrm{N},(4)}$ has the $sp^3$-like distribution, which is similar to those of $\Delta V_i^{\mathrm{N},(4)}$ and $\Delta V_i^{\mathrm{S},(4)}$ with the opposite sign. Moreover, $\Delta V_i^{\mathrm{N},(4)}$ and $\Delta V_i^{\mathrm{S},(4)}$ have similar size. 

 \begin{figure}
\includegraphics[width=1.\linewidth,clip=]{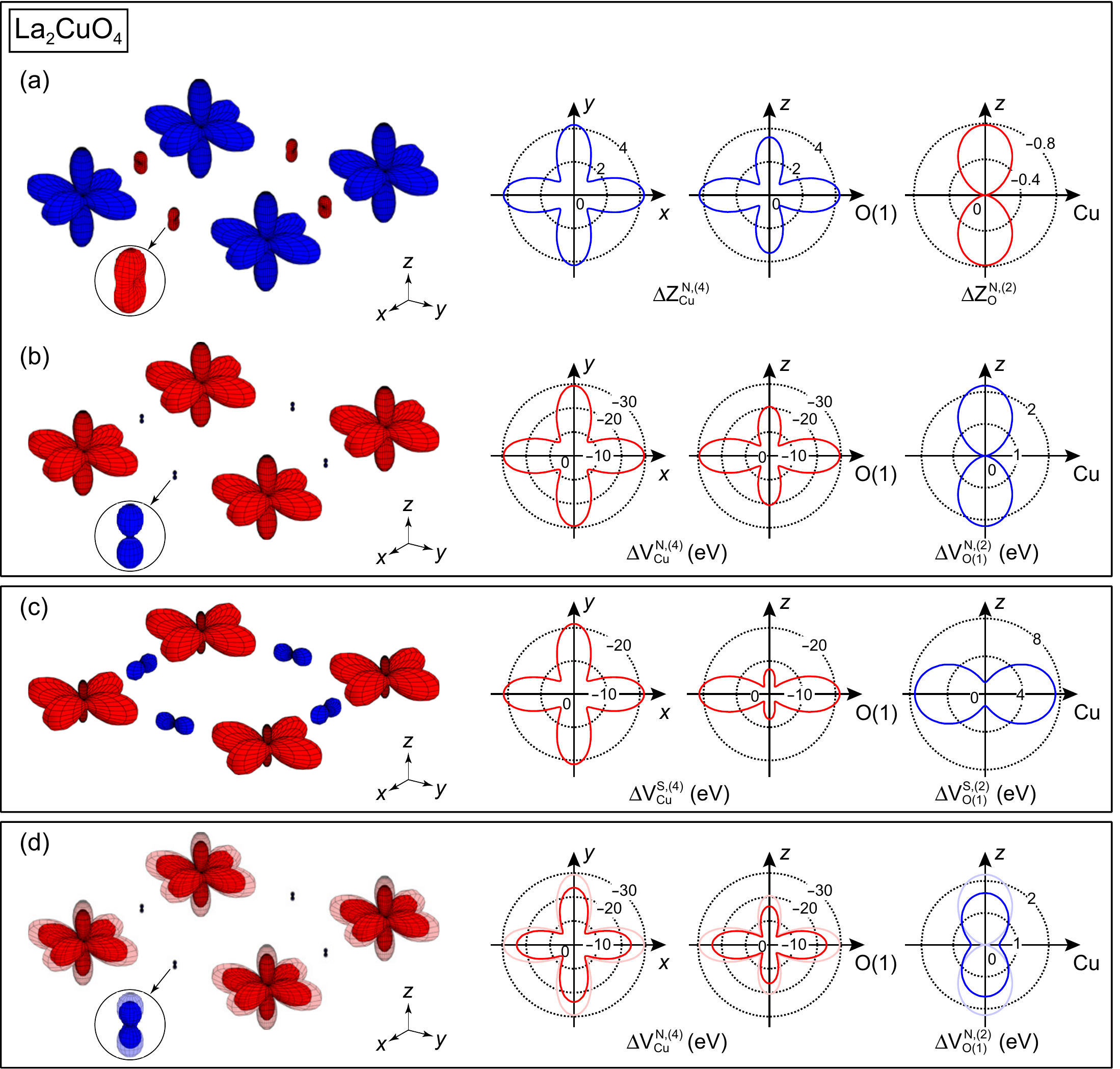}
\caption{\label{fig4} Anisotropic charges and potentials at the Cu and O(1) sites of La$_2$CuO$_4$. (a, b) Non-scalar contributions of the charges (a) and potentials (b). (c) Multipole expansions of the scalar potentials. (d) Non-scalar contributions of the potentials for screened Coulomb interaction ($R_{\mathrm{TF}}=10$ \AA). For each Figure, the left drawing shows the overhead view of the CuO$_2$ plane with the axes of the Cartesian coordinates (In (a), (b) and (d), it includes the close-up view of the non-scalar contribution at the O(1) site). The middle two graphs indicate the anisotropy in the $xy$ and $zb_{\mathrm{CuO}}$ planes at the Cu site ($b_{\mathrm{CuO}}$ is the Cu-O(1) bond direction). The right graph displays the anisotropy in the $zb_{\mathrm{CuO}}$ plane at O(1). The light color plots in (d) indicate the non-scalar potential without screening effects (the same as (b)).}
\end{figure}

There is, nevertheless, no requirement for the similarity between two potentials, $\Delta V_i^{\mathrm{N},(n)}$ and $\Delta V_i^{\mathrm{S},(n)}$, because of the different definitions. For such an example, La$_2$CuO$_4$, a parent material of a hole-doped high-$T_\mathrm{c}$ cuprate ($T_{\mathrm{c},Max}\sim $40 K), is considered. The following tetragonal lattice parameters are used, $a$=3.803\AA, $c$=13.107 \AA, $z$(La)=0.362, and $z$(O(2))=0.184, based on Ref.~\cite{La214struct}. The charge parameters are La$^{3+}$, Cu$^{2+}$, and O$^{2-}$. For this material, the contributions to the 2nd order ($n$=2) are taken into account at the O(1) site in the CuO$_2$ plane, assuming no $sp$-hybridization. The contributions to the 4th order ($n=$4) are taken at the Cu site. Figure 4 shows the non-scalar charges and potentials (Figs.~4(a) and (b)) and the multipole expansion of the scalar potentials (Fig.~4(c)), at the Cu and O(1) sites in the CuO$_2$ plane. 

With respect to two common features mentioned above, one feature is still found in this material; the non-scalar potentials at the Cu and O(1) sites ($\Delta V_{\mathrm{Cu}}^{\mathrm{N},(4)}$ and $\Delta V_{\mathrm{O}(1)}^{\mathrm{N},(2)}$ in Fig.~4(b)) are similar in shape to the respective charges ($\Delta Z_{\mathrm{Cu}}^{\mathrm{N},(4)}$ and $\Delta Z_{\mathrm{O}(1)}^{\mathrm{N},(2)}$ in Fig.~4(a)) with the opposite signs. On the other hand, the other feature depends on the site. The multipole expansion of the scalar potential around the Cu-ion ($\Delta V_{\mathrm{Cu}}^{\mathrm{S},(4)}$, Fig.~4(c)) has similar shape and size to the non-scalar potential ($\Delta V_{\mathrm{Cu}}^{\mathrm{N},(4)}$, Fig.~4(b)) (Note, the $d_{z^2}$(hole)-like component is enhanced in $\Delta V_{\mathrm{Cu}}^{\mathrm{N},(4)}$ more than $\Delta V_{\mathrm{Cu}}^{\mathrm{S},(4)}$, similar to the $p_z$-like component in TiO$_2$ (Fig.~3)). On the other hand, the scalar potential around O(1) ($\Delta V_{\mathrm{O}(1)}^{\mathrm{S},(2)}$, Fig.~4(c)) has $p_x$- or $p_y$-like distribution along the Cu-O(1) bond direction, but the non-scalar potential ($\Delta V_{\mathrm{O}(1)}^{\mathrm{N},(2)}$, Fig.~4(b)) has $p_z$-like distribution perpendicular to the CuO$_2$ plane.

Figure 4(c) indicates that Cu$^{2+}$ site mainly has $d_{x^2-y^2}$(hole)-like component with respect to the scalar potential ($\Delta V_{\mathrm{Cu}}^{\mathrm{S},(4)}$). Given the negative $d_{x^2-y^2}$-like non-scalar contribution at the Cu site ( $\Delta V_{\mathrm{Cu}}^{\mathrm{N},(4)}$, Fig.~4(b)), the energy at this site has negative anisotropy in the Cu-O bond direction. This apparently corresponds to the Heisenberg interaction, describing the antiferromagnetic order in the CuO$_2$ plane. Furthermore, in the (slightly) doped material, the non-scalar potential should be affected by screening; when each site feels (isotropic) Thomas-Fermi screening, the potential is expressed as 
\begin{equation}
V(\mathbf{r}_i)=\sum\limits_{j \ne i}{ \frac{Z_je}{4\pi\varepsilon_0 \left| \mathbf{r}_j - \mathbf{r}_i \right|} \exp \left(-\frac{ \left| \mathbf{r}_j - \mathbf{r}_i \right|}{R_{\mathrm{TF}}}\right)},
\end{equation}
instead of Eq.~(1) ($R_{\mathrm{TF}}$ is the screening length). Figure 4(d) shows the non-scalar potential ($\Delta V_{\mathrm{Cu}}^{\mathrm{N},(4)}$ and $\Delta V_{\mathrm{O}(1)}^{\mathrm{N},(2)}$) with screening of $ R_{\mathrm{TF}}$=10\AA. This decrease of the anisotropic potential with doping suggests decrease of the Heisenberg interaction, that supports the experimental results.

In the discussions above, the screening is isotropic. However, slightly hole-doped La$_2$CuO$_4$ actually shows insulating behavior along the $c$-axis, and the non-scalar potential in the $z$-direction is expected to be less affected than in the $xy$-direction with doping. This may support the superconductivity: the non-scalar potential in the $z$ direction at the O(1) site (Figs.~4(a) and (b)) results in attractive interaction for the hole in the CuO$_2$ plane. This scenario may be supported by the Compton scattering measurements, which reveal that the carrier of underdoped La$_{2-x}$Sr$_x$CuO$_4$ is concentrated around O(1).\cite{Sakurai} 

\begin{figure}
\includegraphics[width=1.\linewidth,clip=]{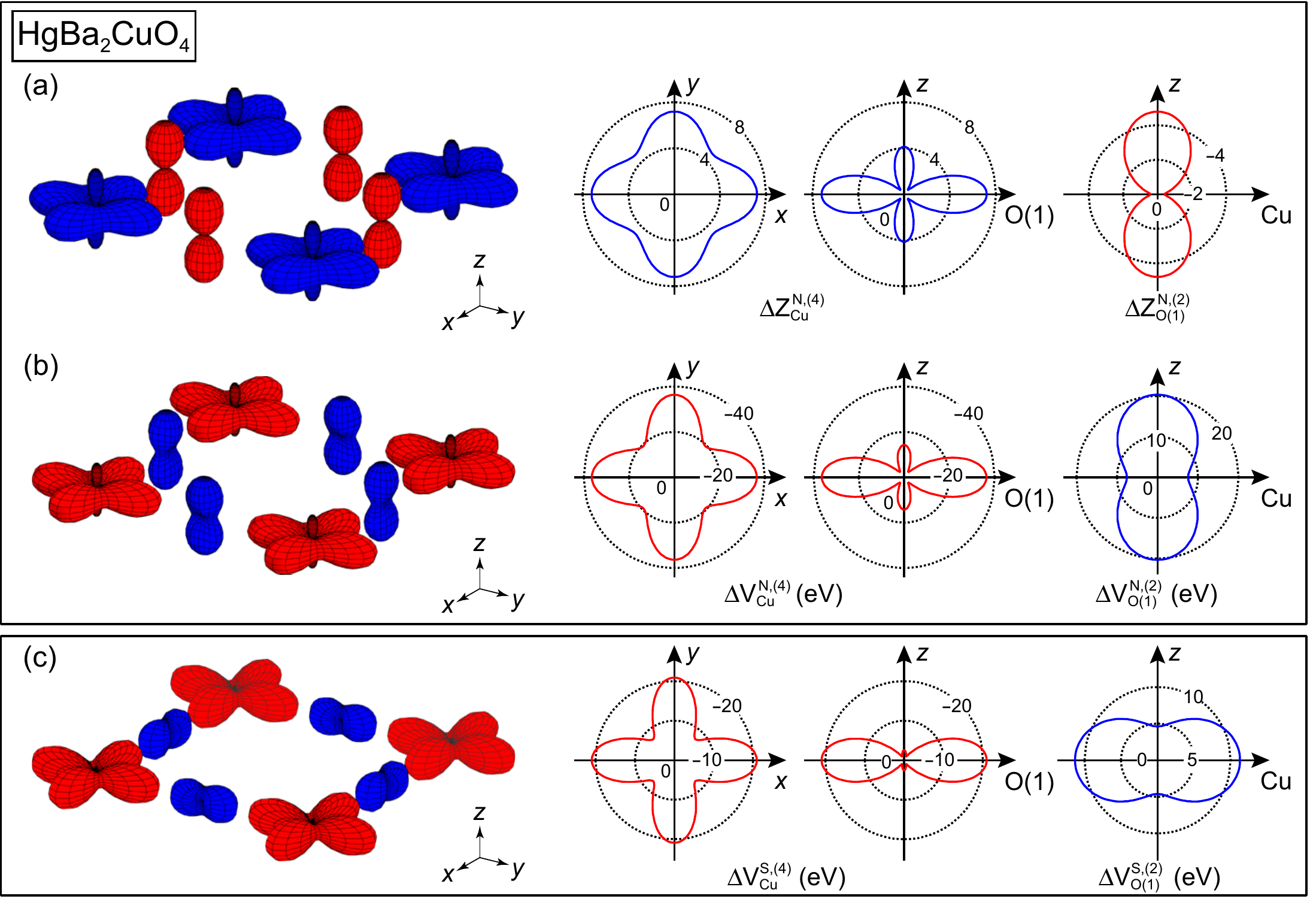}
\caption{\label{fig5} Anisotropic charges and potentials at the Cu and O(1) sites of HgBa$_2$CuO$_4$. (a, b) Non-scalar contributions of the charges (a) and potentials (b). (c) Multipole expansions of the scalar potentials. All drawings and graphs are depicted in the same style as Fig.~4.} 
\end{figure}
 
To confirm this scenario, another parent material of hole-doped high-$T_{\mathrm{c}}$, HgBa$_2$CuO$_4$, is considered. This material can reach higher $T_{\mathrm{c},Max} $ ($\sim$ 95 K) after best doping of holes. Here, the lattice parameters of underdoped HgBa$_2$CuO$_{4+\delta}$ ($T_\mathrm{c}$=59 K) are used; $a$=3.889\AA, $c$=9.540 \AA, $z$(Ba)=0.3016, and $z$(O(2))=0.2061.\cite{Hg1201struct} The charges are given as Hg$^{2+}$, Ba$^{2+}$, Cu$^{2+}$, and O$^{2-}$. Following the style of Fig.~4 for La$_2$CuO$_4$, the anisotropic potentials and charges are displayed in Fig.~5. The non-scalar contributions at O(1) ($\Delta Z_{\mathrm{O}(1)}^{\mathrm{N},(2)}$ and $\Delta V_{\mathrm{O}(1)}^{\mathrm{N},(2)}$, Figs.~5(a) and (b)) is much larger than those in La$_2$CuO$_4$ (Figs.~4(a) and (b)). This larger anisotropy may stabilize the hole along the Cu-O(1) direction in the doped material more, leading to higher $T_{\mathrm{c}}$. Furthermore, each $d_{z^2}$(hole)-like component of the potentials and charge at the Cu site ($\Delta Z_{\mathrm{Cu}}^{\mathrm{N},(4)}$, $\Delta V_{\mathrm{Cu}}^{\mathrm{N},(4)}$, and $\Delta V_{\mathrm{Cu}}^{\mathrm{S},(4)}$) in Fig.~5 is suppressed more than that in La$_2$CuO$_4$ (Fig.~4). Similar suppression is also pointed out by the first principle calculations,\cite{Arita} which suggest enhancement of the superconductivity. 

Except the features above, the anisotropic potentials and charges in Fig.~5 show good agreement with those in Fig.~4 for La$_2$CuO$_4$. The non-scalar potentials at Cu and O(1) have similar symmetries to the respective charges with the opposite signs (Figs.~5(a) and (b)). Moreover, the $d_{z^2}$(hole)-like component around the Cu site estimated by the scalar potential (Fig.~5(c)) is suppressed more than those by the non-scalar potential and charge (Figs.~5(a) and (b)), similarly observed in TiO$_2$ and La$_2$CuO$_4$.

High-$T_{\mathrm{c}}$ cuprates have another type of superconductivity, or superconductivity by electron doping. In order to discuss the anisotropic potentials and charges for electron-doped high-$T_{\mathrm{c}}$ cuprates, a parent material, Nd$_2$CuO$_4$, is considered. The lattice and charge parameters are given as $a$=3.945\AA, $c$=12.176 \AA, $z$(Nd)=0.6489, Nd$^{3+}$, Cu$^{2+}$, and O$^{2-}$,\cite{Nd214struct} and the results are shown in Fig.~6. The style of this Figure is the same as Figs.~4 and 5 for La$_2$CuO$_4$ and HgBa$_2$CuO$_4$. Even in this material, the non-scalar potentials at Cu and O(1) ($\Delta V_{\mathrm{Cu}}^{\mathrm{N},(4)}$ and $\Delta V_{\mathrm{O}(1)}^{\mathrm{N},(2)}$ in Fig.~6(b)) are similar in shape to the respective charges ($\Delta Z_{\mathrm{Cu}}^{\mathrm{N},(4)}$ and $\Delta Z_{\mathrm{O}(1)}^{\mathrm{N},(2)}$ in Fig.~6(a)) with the opposite signs. Additionally, the non-scalar potential at the Cu site ($\Delta V_{\mathrm{Cu}}^{\mathrm{N},(4)}$) is similar in shape and size to the scalar potential ($\Delta V_{\mathrm{Cu}}^{\mathrm{S},(4)}$), except for the suppression of the $d_{z^2}$(hole)-like component in $\Delta V_{\mathrm{Cu}}^{\mathrm{S},(4)}$.

 \begin{figure}
\includegraphics[width=1.\linewidth,clip=]{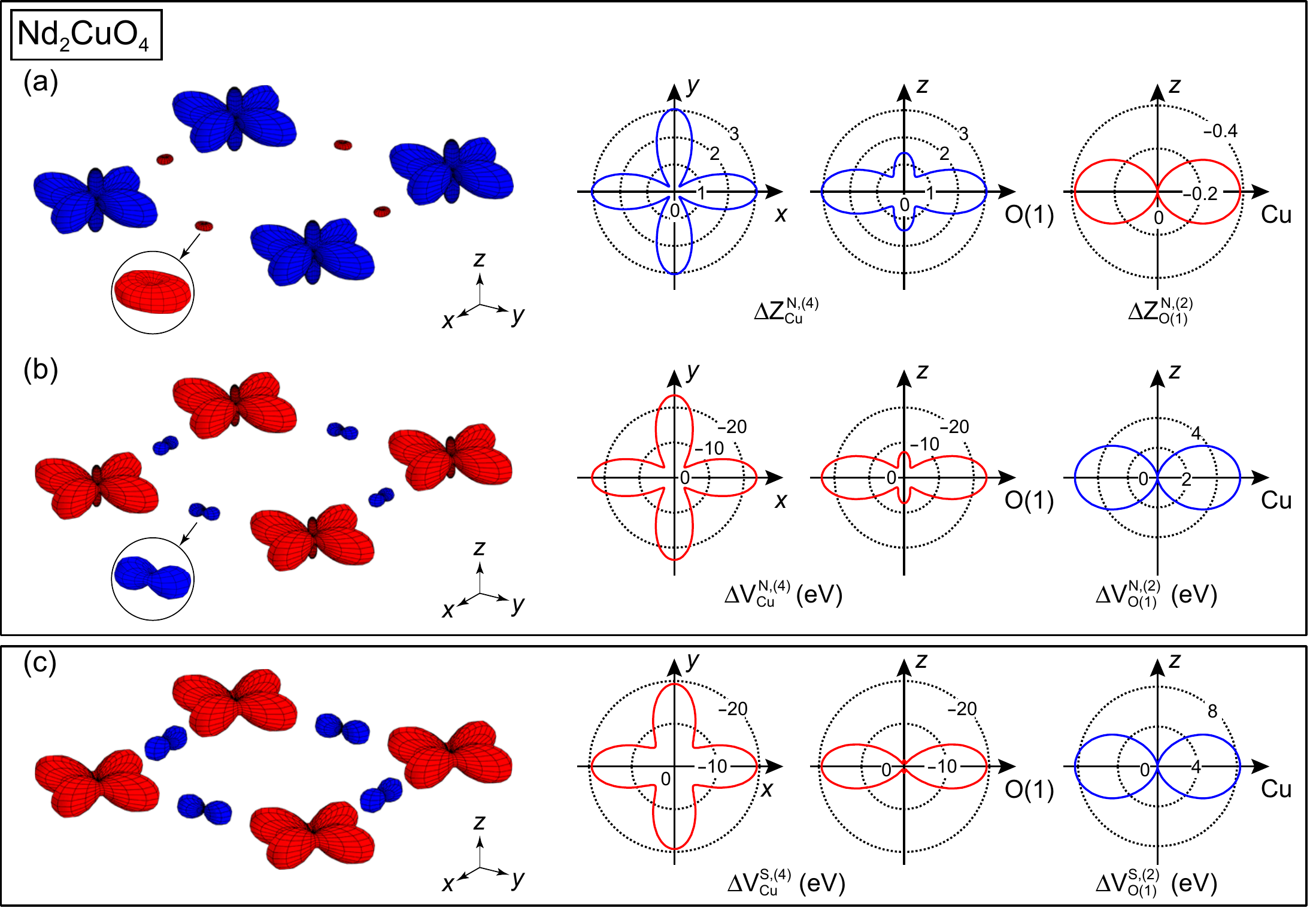}
\caption{\label{fig6} Anisotropic charges and potentials at the Cu and O(1) sites of Nd$_2$CuO$_4$. (a, b) Non-scalar contributions of the charges (a) and potentials (b). (c) Multipole expansions of the scalar potentials. All drawings and graphs are depicted in the same style as Figs.~4 and 5.}
\end{figure}

In contrast to La$_2$CuO$_4$ and HgBa$_2$CuO$_4$, however, the non-scalar potential at the O(1) site ($\Delta V_{\mathrm{O}(1)}^{\mathrm{N},(2)}$ in Fig.~6(b)) is similar to the scalar potential around O(1) ($\Delta V_{\mathrm{O}(1)}^{\mathrm{S},(2)}$, Fig.~6(c)). This similarity rather agrees with SrTiO$_3$ (Fig.~2), TiO$_2$ (Fig.~3), and the other materials with simpler structures. It is supposed not to assist the hole-doped superconductivity, because no attraction occurs at the O(1) site. It should be noted that, each $d_{z^2}$(hole)-like component of $\Delta Z_{\mathrm{Cu}}^{\mathrm{N},(4)}$, $\Delta V_{\mathrm{Cu}}^{\mathrm{N},(4)}$, and $\Delta V_{\mathrm{Cu}}^{\mathrm{S},(4)}$ is suppressed more than that in La$_2$CuO$_4$ (Fig.~4). This suppression is similar to HgBa$_2$CuO$_4$ (Fig.~5), and may be correlated to the superconductivity.

In this manuscript, the discussions are limited in the ionic model of the classical theory in order to highlight the existence of the non-scalar contributions of potential and charge; the existence indicates that an interaction beyond the scalar potential field exists in the ionic (and slightly doped) materials. In other words, there is a possibility of materials which have (anisotropic) exchange interaction and dielectric constant beyond band theory. The intrinsic difference between the scalar and non-scalar potentials may be correlated to difference between the magnetic and orbital degrees of freedom. At the same time, due to the simple picture, further contributions beyond the point-charge model --for example, charge density, charge transfer, quantum spin, and so on-- are ignored in this manuscript. For example, the charge density obtained by the scalar potential will affect the non-scalar potential, as well as the rotational symmetry at the site. In the quantum theory, the non-scalar potential is supposed to modify the original wavefunctions. Further investigation is required to estimate detailed correlation between these contributions.

\section{Conclusions}
To summarize, in the presence of the long-range Coulomb interaction, the potential has two different contributions, which are given by Legendre expansions. One is the conventional scalar contribution, which is expressed as a multipole expansion. The other is the non-scalar contribution, which is accompanied by the non-scalar charge. The non-scalar potential does not affect the scalar contribution, but is correlated to the classical Heisenberg interaction. The scalar and non-scalar potentials have absolute convergence in the Ewald method, and have similar shapes and sizes for most of the ions despite the different definitions. Consequently, in these materials, the non-scalar potentials are expected to give additional effects to the anisotropic states caused by the scalar potentials. On the other hand, the potentials have different symmetries in the parent materials of hole-doping high-$T_{\mathrm{c}}$ cuprates, La$_2$CuO$_4$ and HgBa$_2$CuO$_4$. This difference may cause the superconductivity. Estimation of the non-scalar potential is expected to be important to understand physical properties of ionic and slightly-doped materials.

\section*{Acknowledgments}
The author thanks R. Arita for helpful discussions.

\end{document}